\documentclass[12pt]{article}
\usepackage{epsfig}
\begin{document}
\begin{center}
{\Large {\bf Upper bound on the cutoff in lattice Electroweak theory}


{
\vspace{1cm}
{ A.I.Veselov, M.A.~Zubkov }\\
\vspace{.5cm} {\it  ITEP, B.Cheremushkinskaya 25, Moscow, 117259, Russia }}}
\end{center}

\begin{abstract}
We investigate numerically lattice Weinberg - Salam model without fermions for
realistic values of the fine structure constant and the Weinberg angle. We also
analyze the data of the previous numerical investigations of lattice
Electroweak theory. We have found that moving along the line of constant
physics when the lattice spacing $a$ is decreased, one should leave the
physical Higgs phase of the theory at a certain value of $a$. Our estimate of
the minimal value of the lattice spacing is $a_c = [430\pm 40 {\rm Gev}]^{-1}$.
\end{abstract}

\section{Introduction}

In this paper we consider lattice realization of Electroweak theory (without
fermions). The phase diagram of the correspondent lattice model contains
physical Higgs phase, where scalar field is condensed and gauge bosons $Z$ and
$W$ acquire their masses. This physical phase is bounded by the phase
transition surface. Crossing this surface one leaves the physical phase and
enters the phase of the lattice theory that has nothing to do with the
conventional continuum Electroweak theory. In the physical phase of the theory
the Electroweak symmetry is broken spontaneously while in the unphysical phase
the  Electroweak symmetry is not broken. Thus the unphysical phase is called
also symmetric phase while the Higgs phase is called broken phase of the
theory.

In lattice theory the ultraviolet cutoff is finite and is equal to the inverse
lattice spacing: $\Lambda = \frac{1}{a}$, where $a$ is the lattice spacing.
Alternatively, the Ultraviolet cutoff in lattice theory can be defined as the
momentum $\tilde{\Lambda} = \frac{\pi}{a}$ (see, for example,  \cite{UV}).
Later we shall imply the first definition of the cutoff.

The physical scale can be fixed, for example, using the value of the $Z$-boson
mass $M^{\rm phys}_Z \sim 90$ GeV. Therefore the lattice spacing is evaluated
to be $a \sim [90\,{\rm GeV}]^{-1} M_Z$, where $M_Z$ is the $Z$ boson mass in
lattice units. Within the physical phase of the theory the lines of constant
physics (LCP) are defined that correspond to constant renormalized physical
couplings (the fine structure constant $\alpha$, the Weinberg angle $\theta_W$,
and Higgs mass to Z-boson mass ratio $\eta = M_H/M_Z$). The points on LCP are
parametrized by the lattice spacing. In general, there are two possibilities:
either LCP correspondent to realistic values of $\alpha$, $\theta_W$, and
$\eta$, remains inside the given phase when $a$ is decreased, or it crosses the
boundary at a certain value of $a = a_c$. In the second case $\Lambda_c =
\frac{1}{a_c}$ is the maximal possible ultraviolet cutoff in the lattice
Electroweak theory.

We investigate numerically lattice realization of Weinberg - Salam model. Also
we analyze existing data of the numerical investigation of the $SU(2)$ Gauge -
Higgs model. We find the indications that  there exists the maximal possible
ultraviolet cutoff $\Lambda_c$. Our estimate is $\Lambda_c = \frac{1}{a_c} =
 430\pm 40$ Gev. (With the definition $\tilde{\Lambda}_c = \frac{\pi}{a_c}$ we
arrive at the value of the cutoff $\tilde{\Lambda}_c = 430 \pi \sim 1.3$ Tev.)
It is important to compare this result with the limitations on the Ultraviolet
Cutoff, that come from the perturbation theory.

First, from the point of view of perturbation theory the energy scale $1$ TeV
appears in the Hierarchy problem \cite{TEV}. Namely, the mass parameter $\mu^2$
for the scalar field receives a quadratically divergent contribution in one
loop. Therefore, the initial mass parameter ($\mu^2= - \lambda_c v^2$, where
$v$ is the vacuum average of the scalar field) should be set to infinity in
such a way that the renormalized mass $\mu^2_R$ remains negative and finite.
This is the content of the so-called fine tuning. It is commonly believed that
this fine tuning is not natural \cite{TEV} and, therefore, one should set up
the finite ultraviolet cutoff $\Lambda$. From the requirement that the one-loop
contribution to $\mu^2$ is less than $10 |\mu_R^2|$ one derives that $\Lambda
\sim 1$ TeV. However, strictly speaking, the possibility that the mentioned
fine tuning takes place is not excluded.

In the perturbation theory there is also more solid limitation on the
Ultraviolet cutoff. It appears as a consequence of the triviality problem,
which is related to Landau pole in scalar field self coupling $\lambda$ and in
the fine structure constant $\alpha$. The Landau pole in fine structure
constant is related to the fermion loops and, therefore, has no direct
connection with our lattice result (we neglect dynamical fermions in our
consideration). Due to the Landau pole the renormalized $\lambda$ is zero, and
the only way to keep it equal to its measured value is to impose the limitation
on the cutoff. That's why the Electroweak theory is usually thought of as a
finite cutoff theory. For small Higgs masses (less than about $350$ Gev) the
correspondent energy scale calculated within the perturbation theory is much
larger, than $1$ Tev. The consideration, however, becomes nontrivial when
$\lambda \rightarrow \infty$, and the perturbation expansion in $\lambda$
cannot be used. In this case Higgs mass approaches its absolute upper
bound\footnote{According to the previous investigations of the $SU(2)$ Gauge -
Higgs model this upper bound cannot exceed $10 M_W$.}, and both triviality and
Hierarchy scales approach each other.

\section{Lattice Weinberg - Salam model}

Below we use the following lattice variables:

1. The gauge field ${\cal U} = (U, \theta)$, where
\begin{eqnarray}
 \quad U = \left( \begin{array}{c c}
 U^{11} & U^{12}  \\
 -[U^{12}]^* & [U^{11}]^*
 \end{array}\right)
 \in SU(2), \quad e^{i\theta} \in U(1),
\end{eqnarray}
realized as link variables.

2. A scalar doublet
\begin{equation}
 \Phi_{\alpha}, \;\alpha = 1,2.
\end{equation}

The  action can be considered in the following form
\begin{eqnarray}
 S & = & \beta \!\! \sum_{\rm plaquettes}\!\!
 ((1-\mbox{${\small \frac{1}{2}}$} \, {\rm Tr}\, U_p )
 + \frac{1}{{\rm tg}^2 \theta_W} (1-\cos \theta_p))+\nonumber\\
 && - \gamma \sum_{xy} Re(\Phi^+U_{xy} e^{i\theta_{xy}}\Phi) + \sum_x (|\Phi_x|^2 +
 \lambda(|\Phi_x|^2-1)^2), \label{S}
\end{eqnarray}
where the plaquette variables are defined as $U_p = U_{xy} U_{yz} U_{wz}^*
U_{xw}^*$, and $\theta_p = \theta_{xy} + \theta_{yz} - \theta_{wz} -
\theta_{xw}$ for the plaquette composed of the vertices $x,y,z,w$. Here
$\lambda$ is the scalar self coupling, and $\gamma = 2\kappa$, where $\kappa$
corresponds to the constant used in the investigations of the $SU(2)$ gauge
Higgs model. $\theta_W$ is the Weinberg angle. Bare fine structure
 constant $\alpha$ is expressed through $\beta$ and $\theta_W$ as
\begin{equation}
\alpha = \frac{{\rm tg}^2 \theta_W}{\pi \beta(1+{\rm tg}^2
\theta_W)}.\label{alpha}
\end{equation}

 The renormalized Weinberg
angle is to be calculated through the ratio of the lattice masses:  ${\rm cos}
\, \theta_W = M_W/M_Z$. The renormalized fine structure constant can be
extracted through the potential for the infinitely heavy external charged
particles.

Lattice model with the action (\ref{S}) was investigated numerically in the
number of papers. Most of the papers dealt with the $SU(2)$ Gauge - Higgs
model, i.e. with the case $\theta_W=\pi/2$. The system with arbitrary
$\theta_W$ has been investigated numerically at unphysically large $\alpha$ in
\cite{SU2U1}. Here we list some of the papers that consider $SU(2)$ Gauge -
Higgs model at realistic values of $\beta$ around $\beta = 8$:
\cite{1,2,3,4,5,6,7,8,9,10,11,12,13,14}. Implying that the hypercharge field is
to be included into consideration perturbatively, one can use  expression
(\ref{alpha}) with ${\rm sin}^2 \theta_W = 0.23$ and estimate $\alpha =
\frac{1}{110}$ that is not far from its physical value $\alpha(M_W) =
\frac{1}{128}$.

\section{Numerical investigation of the model at $\theta_W=\pi/6$}

Here we report the results of our numerical investigation of the system
(\ref{S})  for $\theta_W = \frac{\pi}{6}$ (corresponds to ${\rm sin}^2 \theta_W
= 0.25$), $\lambda \rightarrow \infty$, and renormalized $\alpha$ around
$\alpha(M_W) = \frac{1}{128}$. From the very beginning we fix the unitary gauge
$\Phi_1 = const.$, $\Phi_2 = 0$.

The following variables are considered as creating a $Z$ boson and a $W$ boson,
respectively:
\begin{eqnarray}
  Z_{xy} & = & Z^{\mu}_{x} \;
 = {\rm sin} \,[{\rm Arg} U_{xy}^{11} + \theta_{xy}],
\nonumber\\
 W_{xy} & = & W^{\mu}_{x} \,= \,U_{xy}^{12} e^{-i\theta_{xy}}.\label{Z1}
\end{eqnarray}
Here, $\mu$ represents the direction $(xy)$.

After fixing the unitary gauge the electromagnetic $U(1)$ symmetry remains:
\begin{eqnarray}
 U_{xy} & \rightarrow & g^\dag_x U_{xy} g_y, \nonumber\\
 \theta_{xy} & \rightarrow & \theta_{xy} -  \alpha_y/2 + \alpha_x/2,
\end{eqnarray}
where $g_x = {\rm diag} (e^{i\alpha_x/2},e^{-i\alpha_x/2})$. There exists a
$U(1)$ lattice gauge field, which is defined as
\begin{equation}
 A_{xy}  =  A^{\mu}_{x} \;
 = \,[-{\rm Arg} U_{xy}^{11} + \theta_{xy}]  \,{\rm mod} \,2\pi
\label{A}
\end{equation}
that transforms as $A_{xy}  \rightarrow  A_{xy} - \alpha_y + \alpha_x$. The
field $W$ transforms as $W_{xy}  \rightarrow  W_{xy}e^{-i\alpha_x}$.

 The $W$ boson field is charged with respect to the $U(1)$
symmetry. Therefore we fix the lattice Landau gauge in order to investigate the
$W$ boson propagator. The lattice Landau gauge is fixed via minimizing (with
respect to the $U(1)$ gauge transformations) the following functional:
\begin{equation}
 F  =  \sum_{xy}(1 - \cos(A_{xy})).
\end{equation}
Then we extract the mass of the $W$ boson from the correlator
\begin{equation}
\frac{1}{N^6} \sum_{\bar{x},\bar{y}} \langle \sum_{\mu} W^{\mu}_{x}
(W^{\mu}_{y})^{\dagger} \rangle   \sim
  e^{-M_{W}|x_0-y_0|}+ e^{-M_{W}(L - |x_0-y_0|)}
\label{corW}
\end{equation}
Here the summation $\sum_{\bar{x},\bar{y}}$ is over the three ``space"
components of the four - vectors $x$ and $y$ while $x_0, y_0$ denote their
``time" components. $N$ is the lattice length in "space" direction. $L$ is the
lattice length in the "time" direction.

The $Z$-boson mass is calculated using the correlator
\begin{equation}
\frac{1}{N^6} \sum_{\bar{x},\bar{y}} \langle \sum_{\mu} Z^{\mu}_{x} Z^{\mu}_{y}
\rangle   \sim
  e^{-M_{Z}|x_0-y_0|}+ e^{-M_{Z}(L - |x_0-y_0|)}
\label{corZ}
\end{equation}

It is worth mentioning, that in the $Z$ - boson channel many photon state also
exists. The mass of the correspondent state on the finite lattice we used is,
however, larger than that of the $Z$ - boson itself.  For example, on the
lattice $16^3\times 24$ the minimal mass of the $3$ - photon state is $M_{3
\gamma} = 2\frac{2\pi}{16}+\frac{4\pi}{16} \sim 1.5$. Moreover, from the point
of view of perturbation theory this state appears in the correlator
(\ref{corZ}) through the virtual loop and is suppressed by the factor
$\alpha^3$.

In order to evaluate the mass of the Higgs boson we use the correlator
\cite{Montvay}:
\begin{equation}
  \sum_{\bar{x},\bar{y}}\langle H_{x} H_{y}\rangle
   \sim
  e^{-M_{H}|x_0-y_0|}+ e^{-M_{H}(L - |x_0-y_0|)} + const,
\label{cor}
\end{equation}
and the following operators that create Higgs bosons:
\begin{equation}
H_V^x = \sum_{y} Re(U^{11}_{xy} e^{i\theta_{xy}});\,H^x_W = \sum_{y}
|W_{xy}|^2; \,H^x_Z = \sum_{y} Z^2_{xy} \label{HW}
\end{equation}

Here $H_V^x, H_W^x, H_Z^x$ are defined at the site $x$, the sum $\sum_y$ is
over its neighboring sites $y$.

We perform the calculation of renormalized fine structure constant $\alpha_R$
using the potential for infinitely heavy external fermions. We consider Wilson
loops for the right-handed external leptons:
\begin{equation}
 {\cal W}^{\rm R}_{\rm lept}(l)  =
 \langle {\rm Re} \,\Pi_{(xy) \in l} e^{2i\theta_{xy}}\rangle.
\label{WR}
\end{equation}
Here $l$ denotes a closed contour on the lattice. We consider the following
quantity constructed from the rectangular Wilson loop of size $r\times t$:
\begin{equation}
 {\cal V}(r) = {\rm log}\, \lim_{t \rightarrow \infty}
 \frac{  {\cal W}(r\times t)}{{\cal W}(r\times (t+1))}.
\end{equation}
Due to exchange by virtual photons at large enough distances we expect the
appearance of the Coulomb interaction
\begin{equation}
 {\cal V}(r) = -\frac{\alpha_R}{r} + const. \label{V1}
\end{equation}
It should be mentioned here, that in order to extract the renormalized value of
$\alpha$ one may apply to $\cal V$ the fit obtained using the Coulomb
interaction in momentum space. The lattice Fourier transform then gives

\begin{eqnarray}
 {\cal V}(r) & = & -\alpha_R \, {\cal U}(r)+ const,\,
\nonumber\\
{\cal U}(r) & = & \frac{ \pi}{N^3}\sum_{\bar{p}\ne 0} \frac{e^{i p_3 r}}{{\rm
sin}^2 p_1/2 + {\rm sin}^2 p_2/2 + {\rm sin}^2
 p_3/2}
 \label{V2}
\end{eqnarray}
Here $N$ is the lattice size, $p_i = \frac{2\pi}{L} k_i, k_i = 0, ..., L-1$. On
large enough lattices at $r << L$ both definitions approach each other. For
example, for $L = 75, r \in [1,10]$ the linear fit to the dependence ${\cal
U}(r)$  on $\frac{1}{r}$ gives ${\cal U}(r) \sim 0.97/r - 0.18$. However, on
the lattices of sizes we used the difference is important. Say, on the lattice
$16^3$ the fit is ${\cal U}(r) \sim 0.71/r - 0.4$ (for $r\in [1,5]$). Thus, the
values of the renormalized $\alpha_R$ extracted from  (\ref{V1}) and (\ref{V2})
are essentially different from each other. Any of the two ways, (\ref{V1}) or
(\ref{V2}), may be considered as the {\it definition} of the renormalized
$\alpha$ on the finite lattice. And there is no particular reason to prefer the
potential defined using the lattice Fourier transform of the Coulomb law in
momentum space. Actually, our study shows that the single $1/r$ fit
approximates $\cal V$ much better. Therefore, we used it to extract $\alpha_R$.
This should be compared with the results of \cite{14}, where for similar
reasons the single $e^{-\mu r}/r$ fit (instead of the lattice Yukawa fit) was
used in order to determine the renormalized coupling constant in the $SU(2)$
Gauge Higgs model.

In Fig. $1$ we present the phase diagram for the lattice model in the $\beta$ -
$\gamma$ plane. Mainly we used lattices of sizes $16^4$. Some results were
checked on the lattices of size $24^4$. For the evaluation of masses we used
lattices $6^3\times 12$, $8^3\times 16$, $12^3\times 24$, and $16^3 \times 24$.
At small values of $\beta$ this system was considered in \cite{SU2U1}. The
dotted vertical line on the left side of the figure represents the
deconfinement phase transition corresponding to the $U(1)$ constituents of the
model. The continuous line corresponds to the transition between the broken and
the symmetric phases of the model. Physical Higgs phase of the system  is
situated in the right upper corner of Fig.~$1$.

The dotted vertical line on the right-hand side of the diagram represents  the
line, where the renormalized $\alpha$ (calculated on the lattice $16^4$) is
constant and is close to its physical value $\frac{1}{128}$. Actually, on the
tree level this would be the straight line $\beta =  \frac{{\rm tg}^2
\theta_W}{\pi \alpha (1+{\rm tg}^2\theta_W )} \sim 10$. According to our
numerical results on the lattice $16^4$ at $\gamma = 1$ and $\beta$ close to
$\beta = 15$ the renormalized $\alpha_R$ is equal to $\frac{1}{128\pm 1}$. In
addition we checked our results on the renormalized $\alpha_R$ on the lattice
$12^3\times 24$. We have found on this lattice the same value of $\alpha_R$
(within the statistical errors) as on the lattice $16^4$. So, we conclude that
the renormalized fine structure constant calculated using our choice of the fit
for the potential is not sensitive to the change of the lattice size. The given
line of constant renormalized $\alpha_R$ is almost the straight line that is
defined by the two points: $[\gamma = 1; \beta = 15]$ and $[\gamma = 1.5; \beta
= 14.81]$. The accuracy of the calculation of $\alpha_R$ is around $1\%$.

The position of the phase transition lines on this figure was localized,
mainly, using methods developed in \cite{BVZ, BVZ1}. In particular, we
considered the behavior of various monopole - like topological defects that
exist in the given model. (For the definition of the correspondent monopole
currents, their density and percolation probability, see \cite{BVZ}.) The
densities and percolation probabilities of the constructed monopole currents
appear to be very sensitive to the phase transitions. Say, the monopole
currents constructed of the field $\theta$ feel the deconfinement phase
transition corresponding to the $U(1)$ constituents of the model. Their
worldlines are extracted from the hypercharge field $\theta$ in the following
way:
\begin{equation}
 j_Y = \frac{1}{2\pi} {}^*d([d 2\theta]{\rm mod}2\pi)
\label{Hyp}
\end{equation}
(Here we used the notations of differential forms on the lattice. For their
definition see \cite{BVZ, BVZ1} and references therein.) The monopole density
is defined as
\begin{equation}
 \rho = \left\langle \frac{\sum_{\rm links}|j_{\rm link}|}{4L\times N^3}
 \right\rangle,
\label{rho}
\end{equation}
where $N$ is the lattice size in "space" direction, $L$ is the lattice size in
"time" direction in lattice units. (We often used asymmetric lattices for the
calculation of the variables related to the monopole properties.) The density
of hypercharge monopoles is nonzero within the confinement - like phase and
falls sharply within the deconfinement phase.  The average action of the model
appears to be inhomogeneous in the small vicinity of the phase transition line.

 The
monopole currents constructed of the field $A$ (in a way similar to
(\ref{Hyp})) feel the transition between the broken and the symmetric phases of
the model: $ j_A = \frac{1}{2\pi} {}^*d([d A]{\rm mod}2\pi) $.
 Their density drops in the physical Higgs phase.
In order to investigate topological defects extracted from the $Z$ - boson
field we use the definition of the $Z$ - boson creation operator different from
(\ref{Z1}):
\begin{equation}
  Z^{'}_{xy}  =  [{\rm Arg} U_{xy}^{11} + \theta_{xy}]\,{\rm mod} \, 2\pi,
\end{equation}
Then we investigate monopole currents constructed of the field $Z^{'}$: $j_Z =
\frac{1}{2\pi} {}^*d([d Z^{'}]{\rm mod}2\pi)$. Their density also drops in the
physical Higgs phase.

In order to localize the position of this transition we also use the
susceptibility $\chi = \langle H_Z^2 \rangle - \langle H_Z\rangle^2$.  In Fig.
$2$ the dependence of the susceptibility on $\gamma$ on the lattice $8^3\times
16$ is represented at fixed $\beta = 15$. $H_Z$ is composed of $Z$ field
according to expression (\ref{HW}). We also check our data represented on Fig.2
using the lattices $12^3\times 24$ and $16^4$. We do not find any dependence of
$\chi$ on the lattice size.

It can be seen that the maximum of the susceptibility composed of $H_Z$
corresponds to the values of $\gamma$ around $\gamma = 0.92$. We found that the
percolation probabilities of both monopole currents extracted from the fields
$A$ and $Z^{'}$ vanish at the same value of $\gamma$. In summary, we evaluate
the position of the transition between the two phases at $\beta = 15$ as
$\gamma_c = 0.92 \pm 0.02$.

It is worth mentioning that according to our numerical results monopoles
extracted from the fields $A$ and $Z^{'}$ are condensed in the unphysical
symmetric phase of the model. The correspondent field configurations carry
magnetic charge and dominate in the vacuum of the symmetric phase. Therefore,
this phase indeed has nothing to do with the real continuum physics.

The behavior of the densities of the considered topological objects is in
general very similar to that of the $SU(2)\times U(1)/Z_2$ model investigated
in \cite{BVZ}. It is worth mentioning that the line of the transition between
the broken and the symmetric phases of the model can actually be a crossover
line. In general we evaluate error bars in determination of the phase
transition points given in Fig.1 as $\Delta \gamma = \pm 0.05; \Delta \beta =
\pm 0.05$ although in some regions of the phase diagram the accuracy is better.

For the calculation of the W-boson and Z-boson masses we used lattices of sizes
$6^3\times 12$, $8^3\times 16$, $12^3\times 24$, and $16^3 \times 24$. It has
been found that the $W$ - boson mass contains an artificial dependence on the
lattice size. We suppose, that this dependence is due to the photon cloud
surrounding the $W$ - boson. The energy of this cloud is related to the
renormalization of the fine structure constant. It has been shown above that
the definition of renormalized $\alpha_R$ is ambiguous on the finite lattice.
The difference between the two possible definitions (via the single $1/r$ fit
and via the lattice Coulomb potential) depends strongly on the lattice size. On
the other hand, the $Z$ - boson correlator does not possess this artificial
dependence on the lattice size. Therefore, we use the $Z$ - boson mass in order
to fix Ultraviolet cutoff in the model.


Careful investigation of the $ZZ$ correlator at the point $\gamma = 1, \beta =
15$ shows that $M_Z$ does not depend on the lattice size. The value of mass
$M_Z = 0.22 \pm 0.01$ at $\gamma = 1, \beta = 15$ was obtained on four
different lattices of sizes $6^3\times 12$, $8^3\times 16$, $12^3\times 24$,
and $16^3 \times 24$. The dependence of the $Z$-boson mass on $\gamma$ at
$\beta = 15$ on the lattice $8^3\times 24$ together with the linear fit are
given in Fig.~$3$. The linear fit is $M_Z = 0.009 + 0.217 \gamma$.

Basing on this data we conclude that the Z - boson mass in lattice units in the
physical Higgs phase of the theory cannot exceed the value $0.21\pm 0.01$ for
$\beta = 15$ as we locate the transition between the two phases at $\gamma =
0.92 \pm 0.02$. At the point [$\beta = 15, \gamma = 0.92$] the value of
renormalized $\alpha_R$ does not deviate much from the value calculated on the
line $\alpha_R = \frac{1}{128}$. Actually, the deviation is within $1\%$. Thus
we expect the maximal possible Ultraviolet cutoff at realistic value of the
fine structure constant cannot exceed $\Lambda_c = 430\pm 40$ Gev \footnote{We
also like to notice here that in the previously investigated $SU(2)$ Gauge -
Higgs model it was found that the gauge boson mass in lattice units grows when
one moves into the physical Higgs phase starting from the transition point
(when the gauge coupling $\beta$ is fixed
\cite{1,2,3,4,5,6,7,8,9,10,11,12,13,14}.)}. So, the Ultraviolet cutoff grows
when $\gamma$ is decreased, its maximal value within the physical Higgs phase
is achieved at the transition point and cannot exceed $\Lambda_c =
\frac{1}{a_c} = 430\pm 40$ Gev (or, $\tilde{\Lambda}_c = \frac{\pi}{a_c} \sim
1.35$ Tev).

As for the Higgs boson mass, due to the insufficient statistics we cannot
extract $M_H$ from our data with reasonable accuracy. According to our (very
rough) estimate at  $\beta=15, \gamma \in [0.8;1.2]$ we have $M_H/M_Z \sim 9
\pm 2$. This estimate is in agreement with the investigation of the $SU(2)$
Gauge Higgs model \cite{12,13,14} performed near the transition point for the
London limit of the Higgs potential and realistic $\beta$. Actually, as in
\cite{12} we made our estimate based on the consideration of the correlator for
small space-time separation ($ \le 3$). It was found in \cite{14} that at
larger distances the second mass parameter close to $2 M_W$ contributes to the
correlator. In \cite{14} in order to evaluate Higgs boson mass in this
situation this second value was considered as the mass of the bound state of
the two gauge bosons, and only the first mass in the given channel was
interpreted as the Higgs boson mass.

\section{The tree level estimates of lattice quantities}

At finite $\lambda$ the line of constant renormalized $\alpha$ is not a line of
constant physics, because the mass of the Higgs boson depends on the position
on this line. Thus, in order to investigate the line of constant physics  one
should vary $\lambda$ together with $\gamma$ to keep the ratio of lattice
masses $M_H/M_W$ constant.

In order to obtain the tree level estimates let us rewrite the lattice action
in an appropriate way. Namely, we define the scalar field $\tilde{\Phi} =
\sqrt{\frac{\gamma}{2}} \Phi$. We have:

\begin{eqnarray}
 S & = & \beta \!\! \sum_{\rm plaquettes}\!\!
 ((1-\mbox{${\small \frac{1}{2}}$} \, {\rm Tr}\, U_p )
 + \frac{1}{{\rm tg}^2 \theta_W} (1-\cos \theta_p))+\nonumber\\
 && + \sum_{xy} |\tilde{\Phi}_x - U_{xy} e^{i\theta_{xy}}\tilde{\Phi}_y|^2 + \sum_x (\mu^2 |\tilde{\Phi}_x|^2 +
 \tilde{\lambda} |\tilde{\Phi}_x|^4) + \omega , \label{S2}
\end{eqnarray}
where $\mu^2 = - 2(4+(2\lambda-1)/\gamma)$, $\tilde{\lambda} =
4\frac{\lambda}{\gamma^2}$, and $\omega = \lambda V$. Here  $V = L^4$ is the
lattice volume, and $L$ is the lattice size.

For negative $\mu^2$ we fix Unitary gauge $\tilde{\Phi}_2=0$, ${\rm Im}\,
\tilde{\Phi}_1 = 0$, and introduce the vacuum value of $\tilde{\Phi}$: $v =
\frac{|\mu|}{\sqrt{2\tilde{\lambda}}}$. We also introduce the scalar field
$\sigma$ instead of $\tilde{\Phi}$: $\tilde{\Phi}_1 = v + \sigma$. We denote
$V_{xy} = (U^{11}_{xy}e^{i\theta_{xy}} - 1)$, and obtain:
\begin{eqnarray}
 S & = & \beta \!\! \sum_{\rm plaquettes}\!\!
 ((1-\mbox{${\small \frac{1}{2}}$} \, {\rm Tr}\, U_p )
 + \frac{1}{{\rm tg}^2 \theta_W} (1-\cos \theta_p))+\nonumber\\
 && + \sum_{xy} ((\sigma_x - \sigma_y)^2 + |V_{xy}|^2 v^2)  + \sum_x 2|\mu|^2 \sigma_x^2 \nonumber\\
 && + \sum_{xy} ((\sigma^2_y+2v \sigma_y)|V_{xy}|^2 - 2(\sigma_x - \sigma_y){\rm Re} V_{xy} (\sigma_y +v) ) + \nonumber\\
 && + \sum_x  \tilde{\lambda} \sigma_x^2 (\sigma_x^2 + 4 v \sigma_x) +  \tilde{\omega} , \label{S2}
\end{eqnarray}
where $\tilde{\omega} = \omega - \tilde{\lambda} v^4 V$.

Now we easily derive the tree level estimates:
\begin{eqnarray}
M_H &=& \sqrt{2}|\mu| = 2\sqrt{4+(2\lambda-1)/\gamma}; \nonumber\\
M_W &=& \sqrt{2} \frac{v}{\sqrt{\beta}} =  \sqrt{\frac{\gamma(4\gamma+2\lambda-1)}{2\lambda\beta}}; \nonumber\\
M_W &=& {\rm cos}\theta_W M_Z\nonumber\\
M_H/M_W &=& \sqrt{8\lambda \beta/\gamma^2};\nonumber\\
\Lambda &=& \sqrt{\frac{2\lambda\beta}{\gamma(4\gamma+2\lambda-1)}} \, [80\,
{\rm GeV}];\label{tree}
\end{eqnarray}
the fine structure constant is given by the formula (\ref{alpha}) and does not
depend on $\lambda$ and $\gamma$. From (\ref{tree}) we learn that at the tree
level LCP on the phase diagram corresponds to fixed $\beta = \frac{{\rm tg}^2
\theta_W}{\pi \alpha(1+{\rm tg}^2 \theta_W)} \sim 10 $ and $\eta = M_H/M_W$,
and is given by the equation $\lambda(\gamma) = \frac{\eta^2}{8\beta}
\gamma^2$. Actually, numerical research shows that the real LCP stays not far
from this tree level estimate (for $\lambda << 1$).

The important case is $\lambda = \infty$, where the tree level estimates give
\begin{eqnarray}
M_H &=& \infty; \nonumber\\
M_W &=& \sqrt{\frac{\gamma}{\beta}}; \nonumber\\
M_Z &=& \sqrt{\frac{\gamma}{\beta}}{\rm cos}^{-1}\theta_W; \nonumber\\
\Lambda &=& \sqrt{\frac{\beta}{\gamma}} \, [80\, {\rm GeV}];\label{treei}
\end{eqnarray}

In the $SU(2)$ gauge Higgs model for the small values of $\lambda << 0.1$ the
tree level estimate for $M_H/M_W$ gives values that differ from the
renormalized ratio by about 20\%\cite{11}.
The tree level estimate for the ultraviolet cutoff is about $310$ GeV at
$\lambda = \infty,\gamma = 1, \beta = 15$ that is not far from the numerical
result given in the previous section. In the $SU(2)$ Gauge Higgs model at
$\lambda = \infty$ the critical $\gamma_c = 0.63$ for $\beta = 8$ \cite{14}. At
 this point the tree level estimate gives $\Lambda = 285$ Gev while the direct measurements
 give $\Lambda \in [270; 470]$ Gev for values of $\gamma \in [0.64; 0.95]$ \cite{14}. The investigations of
the $SU(2)$ Gauge Higgs model showed that a consideration of finite $\lambda$
does not change much the estimate for the gauge boson mass. However, at finite
$\lambda$ and values of $\gamma$ close to the phase-transition point the tree
level formula does not work at all.

The tree level estimate for the critical $\gamma$ is $\gamma_c =
(1-2\lambda)/4$. At small $\lambda$ this formula gives values that are close to
the ones obtained by the numerical simulations \cite{12,13,14}. In particular,
$\gamma_c \rightarrow 0.25$ ($\kappa_c \rightarrow 0.125$) at $\lambda << 1$.
However, this formula clearly does not work for $\lambda > 1/2$. From
\cite{Montvay,12,13,14} we know that the critical coupling in the $SU(2)$ Gauge
Higgs model is about $2 - 4$ times smaller for $\lambda =0$ than for $\lambda =
\infty$.

\section{Analysis of the existing data}

From the previous research we know that the phase diagram in the $\beta$ -
$\gamma$ plane of the lattice $SU(2)$ Gauge - Higgs for any fixed $\lambda$
resembles the phase diagram represented in the figure $1$. The only difference
is that in the $SU(2)$ Gauge - Higgs model the confinement-deconfinement phase
transition corresponding to the $U(1)$ constituents of the model is absent. The
direct measurement of the renormalized coupling $\beta_R$ shows
\cite{1,2,3,4,5,6,7,8,9,10,11,12,13,14} that the line of constant renormalized
coupling constant (with the value close to the experimental one) intersects the
phase transition line. Also we know from the direct measurements of $M_W$ in
the $SU(2)$ Gauge - Higgs model that the ultraviolet cutoff is increased when
one is moving along this line from the physical Higgs phase to symmetric phase.
It is also worth mentioning that the line of the transition between the broken
and the symmetric phases of the model can actually be a crossover line. .

According to (\ref{tree}) the W-boson mass in lattice units vanishes at the
critical $\gamma_c = (1-2\lambda)/4$. This means that the tree level estimate
predicts the appearance of an infinite ultraviolet cutoff at the transition
point for finite $\lambda$. At infinite $\lambda$ the tree level estimate gives
nonzero values of lattice $M_W$ for any nonzero $\gamma$. Our numerical
investigation of $SU(2)\otimes U(1)$ model (at infinite $\lambda$) and previous
calculations in the $SU(2)$ Gauge Higgs model (both at finite $\lambda$ and at
$\lambda = \infty$) showed that for the considered lattice sizes renormalized
masses do not vanish and the transition is either of the first order or a
crossover. (Actually, the situation, when the cutoff tends to infinity at the
position of the transition point means that there is a second order phase
transition.) The dependence on the lattice sizes for the $SU(2)$ Gauge Higgs
model was investigated, for example, in \cite{10}. Namely, for $\beta = 8$,
$\lambda \sim 0.00116$, where $M_H \sim M_W$, the correlation lengths were
evaluated at the critical value $\kappa_c = \gamma_c/2$. For different lattice
sizes (from $12^3\times 28$ to $18^3 \times 36$) no change in correlation
length was observed \cite{10}.

In the table we summarize the data on the ultraviolet cutoff $\frac{1}{a}$
achieved in selected lattice studies of the $SU(2)$ Gauge Higgs model. ($a$ is
the lattice spacing.) Everywhere $\beta$ is around $\beta \sim 8$ and the
renormalized fine structure constant is around $\alpha \sim 1/110$.

Among the papers listed in this table there are results of both finite
temperature and zero temperature studies. However, in the case when the finite
temperature simulations are performed the authors either refer to the analogous
simulations of the zero temperature theory or performed such simulations
directly. This is related to the fact that the only way to set up the scale in
the theory and, correspondingly, to calculate the temperature, is to deal with
the zero temperature model on the symmetric lattice. To be explicit, one should
calculate  lattice spacing  $a$ on the symmetric lattice via calculation of the
gauge boson mass. Then on the asymmetric lattice (with the same values of
couplings as on the symmetric one) the value of temperature is $1/(Na)$ , where
$N$ is the lattice size in time direction. The ultraviolet cutoffs used in the
mentioned lattice studies of the finite temperature theory actually correspond
to the zero temperature models, where these values have been calculated.

\begin{table}
\label{tab.01}
\begin{center}
\begin{tabular}{|c|c|c|}
\hline
{\bf Reference}  & {\bf Ultraviolet Cutoff} $\frac{1}{a}$ (GeV) & {\bf $M_H$} (GeV)\\
\hline
\cite{1}  & 140 (space direction) 570 (time direction) & 80 \\
\hline \cite{2}  & 280 (time direction) & 80 \\
\hline \cite{3}  & 280 & 34 \\
\hline \cite{4}  & 110 & 16 \\
\hline \cite{5}  & 90 (space direction) 350 (time direction) & 34 \\
\hline \cite{6}  & 280 & 48 \\
\hline \cite{7}  & 140 & 35 \\
\hline \cite{8}  & 280 & 20 , 50 \\
\hline \cite{9}  & 190 & 50 \\
\hline \cite{10}  & 260 & 57 - 85 \\
\hline \cite{11}  & 200 - 300 & 47 - 108 \\
\hline \cite{12}  & 400 & 480 \\
\hline \cite{13}  & 330 -  470 & 280 - 720
 \\
\hline \cite{14}  & 250 -  470 &  720
($\lambda =\infty$) \\
\hline
\end{tabular}
\end{center}
\end{table}

Strictly speaking, the above described picture works at infinite (or, high
enough) lattice size. If $T \rightarrow 0$, then one should use lattice with
the time extent $N_T = \frac{1}{Ta} \rightarrow \infty$. That's why the value
of lattice spacing calculated on the ideal infinite symmetric lattice is to be
used in the finite temperature study at small enough temperatures. Our analysis
shows, that the smallest value of $a$ is around $[400 {\rm Gev}]^{-1}$ (see
Section $7$ of the present paper). Our study shows also, that $M_Z$ does not
depend on the lattice size $L$ for $L > 5$. Thus for the time extent of the
asymmetric lattice $N_T > 5$ corresponding to $T < 80$ Gev the value of $a$
calculated on the symmetric lattice can be applied. However, already at the
temperatures of the order of $400$ Gev it is necessary to use lattice with the
time extent $N = \frac{1}{Ta} \sim 1$. Therefore, it is obvious, that at $T
> 80$ Gev the lattice theory suffers from lattice artifacts. At the
temperatures larger, than $400$ Gev, it cannot be applied in principle.

In principle, the effect of lattice artifacts could be partially corrected if
the effective value of lattice spacing is used that is different from that of
calculated on the symmetric lattice. If so, the effective upper bound on the
Ultraviolet cutoff $\Lambda_c = \frac{1}{a_c}$ can be considered as depending
on temperature. However, the discussion of such a dependence is out of the
scope of the present paper.

\section{Triviality problem and the Hierarchy
scale}

The emergence of the triviality problem in lattice theory was considered in a
number of papers (see, for example, \cite{13,14}). According to the common view
on the problem the renormalized $\lambda$ tends to zero when the ultraviolet
cutoff tends to infinity. Thus at the infinite value of the cutoff Higgs sector
becomes trivial (noninteracting). As a result the renormalized ratio $M_H/M_W$
should tend to zero when the cutoff tends to infinity while the other
renormalized couplings ($\alpha$ and $\theta_W$) remain constant. However, at
finite ultraviolet cutoff this ratio may remain far from zero. In the
situation, when the measured Higgs boson mass is larger than the inverse
lattice spacing, we cannot consider the Higgs boson as a real quantum state
existing in the theory. (We do not think, however, that in this situation the
theory looses sense at all.) Thus, when $M_H$ becomes of the order of the
cutoff, it approaches its absolute upper bound. This gives the so-called
triviality upper bound on the Higgs mass allowed in lattice Electroweak theory.
According to the previous investigation of the $SU(2)$ Gauge - Higgs model this
triviality bound is $M_H/M_W < 10$ (see, for example, \cite{13,14}).

Basing on the perturbative treatment of the triviality problem one expects that
in the lattice theory this problem appears as follows. Each Line of Constant
Physics (correspondent to fixed renormalized $\alpha$, $\theta_W$, and
$M_H/M_W$) must be ended at a certain value of the cutoff related to the
triviality problem. Basing on the perturbation theory one may expect, that this
value of the cutoff for $M_H < 350$ Gev is larger, than $10$ Tev (see, for
example, \cite{T} and references therein). If $M_H$ approaches its absolute
upper bound $M^c_H$, then the perturbation theory predicts decrease of the
maximal Ultraviolet cutoff $\Lambda_t$ related to the triviality problem.

The Hierarchy scale is around $1$ Tev. So, if the mentioned above picture is
valid, moving along the Line of Constant Physics at $M_H < 350$ Gev we would
encounter the Hierarchy scale much earlier, than the triviality problem.
However, as it will be explained in the next section, there are indications
that the Line of Constant Physics always stops at the point, where the value of
the ultraviolet cutoff $\Lambda_c$ is at the Hierarchy scale. This means that
within the lattice theory the emergence of the triviality problem is more
complicated, than it was usually thought. We suppose, that both Landau pole in
scalar self coupling, and the Hierarchy problem in perturbation theory, as well
as the appearance of the maximal cutoff $\Lambda_c \sim 1$ Tev in the lattice
theory may actually be the manifestations of the same phenomenon.

It is worth mentioning, that if $M_H \rightarrow M^c_H$, then perturbative
$\Lambda_t$ is decreased and approaches the value of the Higgs mass. So,
$\Lambda_t$ and $\Lambda_c$ approach each other (see Fig. 2 of \cite{T}).

\section{The maximal value of the cutoff}

On the lattice the bare mass parameter in lattice units is $\mu^2 = -
2(4+(2\lambda-1)/\gamma)$. In the lattice theory we reach the point where the
renormalized $\mu_R^2$ becomes positive, if we are moving along the line of
constant $\alpha$, while the ultraviolet cutoff $\Lambda$ is increased. This is
the point of a phase transition between the broken and the symmetric phases of
the model.

The content of the fine tuning in continuum approach is that we set up the
initial parameter $\mu^2$ in such a way that the quadratically divergent
contribution to $\mu_R^2$ is cancelled. This means that $-\mu^2$ should be as
large as ${\rm const}\times \Lambda^2$. In the perturbation theory, in
principle, for any given $\Lambda$ we can choose an appropriate value of
$\mu^2$. Therefore the naive guess would be that on the lattice in order to
increase the cutoff the value of bare lattice $\lambda$ should be increased
(then $-\mu^2 = 2(4+(2\lambda-1)/\gamma)$ is increased). In our simulations we
used the maximal possible value of $\lambda$, i.e. $\lambda = \infty$. And we
have found that the value of the cutoff cannot exceed its maximal value
$\Lambda_c$. At infinite $\lambda$ the tree level estimate gives $\Lambda^{\rm
tree}_c = \sqrt{\frac{\beta}{\gamma_c}}[80\, {\rm Gev}]^{-1}$. If we substitute
$\gamma_c \sim 1$ and $\beta \sim 15$ then the tree level estimate  gives
$\Lambda^{\rm tree}_c \sim 310$ Gev. Our calculations\footnote{In the previous
numerical investigations of lattice Electroweak theory at realistic values of
$\beta$ the $U(1)$ constituent of the model was not taken into account. It was
implied that the hypercharge field is to be taken into account using
perturbation expansion. Thus possible nonperturbative effects were ignored.
However, we see that nonperturbative effects are important for evaluation of
maximal possible Ultraviolet cutoff in lattice Weinberg - Salam model (at
least, at $\lambda \rightarrow \infty$).} gave us value $\Lambda_c =
\frac{1}{a_c} \sim 430\pm 40$  for ${\rm sin}^2 \theta_W = 0.25$, $\alpha_R
\sim \frac{1}{128}$. (Here $a_c$ is the value of lattice spacing.) In the
$SU(2)$ Gauge - Higgs model the maximal reported value of $\Lambda =
\frac{1}{a}$ is $470$ Gev. It is worth mentioning here that the weak coupling
expansion in lattice theory \cite{phi4} gives the prediction that the maximal
possible ultraviolet cutoff is achieved in lattice Electroweak theory at
infinite $\lambda$. The value $470$ Gev was obtained when the $U(1)$
constituent of the model was neglected. Moreover, the fine structure constant
in the correspondent research was around $\frac{1}{110}$. In our research the
$U(1)$ subgroup of the Electroweak gauge group is taken into account  and
$\alpha_R$ is around its physical value $\frac{1}{128}$. That's why we feel it
appropriate to estimate the maximal cutoff in the lattice Electroweak theory
(with dynamical fermions neglected) equal to the value calculated in our work.

Thus basing on our data and on the data of the previous numerical research we
expect that $\Lambda_c$ remains finite at the transition point for any
$\lambda$. If so, then in the lattice theory there is no way to avoid entering
the wrong phase while increasing $\Lambda$ with {\it any} choice of initial
parameters of the model. However, the possibility still remains that the second
order phase transition between the symmetric and the broken phases may appear
at selected exceptional values of the coupling constants. Then at these points
the Ultraviolet cutoff may become infinite.

\section{Conclusions and discussion}

To conclude, in this paper we reported  the results of numerical investigation
of the lattice Weinberg - Salam model at infinite bare scalar self coupling. We
also analyzed results of the previous lattice study of $SU(2)$ Gauge - Higgs
model. Both our results and the previous data indicate that the values of
lattice spacings smaller, than a critical value $a_c$, cannot be achieved in
principle. Basing on the existing data we expect, that $a_c$ is about $[430 \pm
40 {\rm Gev}]^{-1}$.

Our study shows that the susceptibility represented in Fig. \ref{fig.6} does
not depend on the lattice size. This can be considered as the indication, that
the transition between the Higgs phase and the symmetric phase of the model is
the crossover. We also have found, that the percolation of monopole - like
topological defects appears as an order parameter for this transition. That's
why we conclude, that the given transition may belong to the class of the
transitions of the so - called Kertesz type (see, for example, \cite{Kertesz}).

The important question is how the minimal value $a_c$ of the lattice spacing
depends on the details of lattice regularization. In particular, one may
suppose that it could become possible to find out the improved lattice action
that allows to decrease $a_c$. However, this question is out of the scope of
the present paper.

It is the common point of view, that due to the triviality problem the Weinberg
- Salam model should be considered as a finite cutoff theory.  The main result
of our paper is that the value of the maximal cutoff in lattice Electroweak
theory is essentially smaller, than it was thought previously. Namely, we
suppose, that it is about $\Lambda_c = \frac{1}{a_c} \sim 430 \pm 40$ Gev (or,
$\tilde{\Lambda}_c = \frac{\pi}{a_c} \sim 1.3$ Tev). Although we neglect
dynamical fermions and consider the scalar field potential in London limit, we
suppose that the investigation of the theory with the finite value of scalar
self coupling and with dynamical fermions included will not change our estimate
crucially. Thus we expect, that the Weinberg - Salam model can be used only at
the energies ${\cal E} << 1$ Tev. At the same time at the energies approaching
$1$ Tev the other theory should be used \footnote{The perturbative analysis of
the Hierarchy problem usually leads to the same conclusion if the fine tuning
is treated as unnatural and, therefore, unacceptable.}.

The appearance of the upper bound on the cutoff in lattice Electroweak theory
may have important consequences in finite temperature theory. In particular,
one of the scenarios of baryon asymmetry appearance is related to Electroweak
sphalerons. However, the correspondent energy scale $10$ Tev is far above
$\Lambda_c$. Moreover, the lattice Electroweak theory cannot be
applied\footnote{In this estimate we ignore the possible dependence of
$\Lambda_c$ on $T$ that has been mentioned at the end of Section $5$.} at $T
> \Lambda_c \sim 430$ Gev, because time extent of the lattice is evaluated as $N
\sim \frac{1}{a T}$.

This work was partly supported by RFBR grants 08-02-00661, and 07-02-00237,
RFBR-DFG grant 06-02-04010, by Grant for leading scientific schools 679.2008.2,
by Federal Program of the Russian Ministry of Industry, Science and Technology
No 40.052.1.1.1112. The essential part of numerical simulations was done using
the facilities of Moscow Joint Supercomputer Center.

\clearpage

\begin{figure}
\begin{center}
 \epsfig{figure=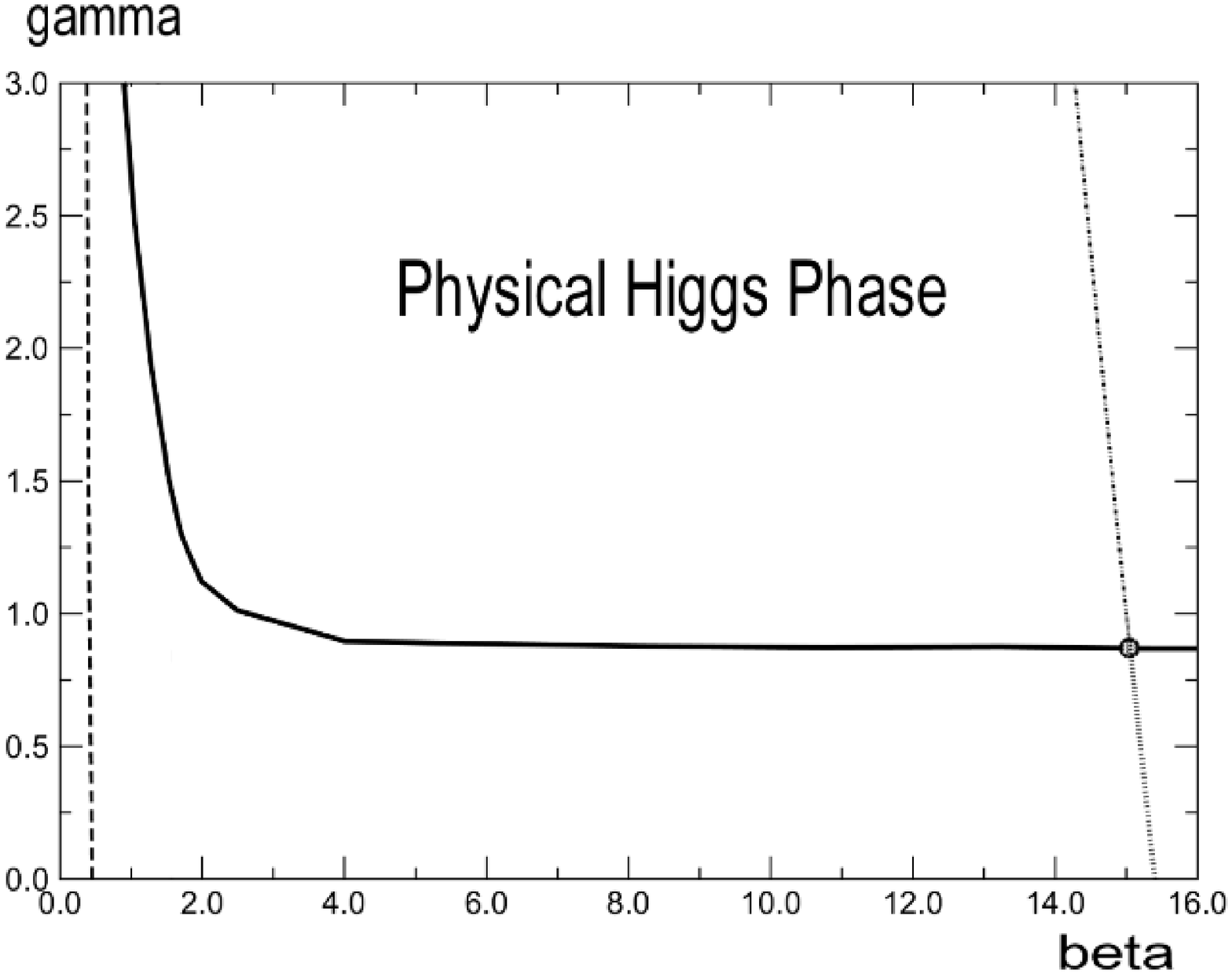,height=60mm,width=80mm,angle=0}
\caption{\label{fig.1} The phase diagram of the lattice model at fixed
$\lambda$ in the
 $(\beta, \gamma)$-plane.}
\end{center}
\end{figure}

\begin{figure}
\begin{center}
 \epsfig{figure=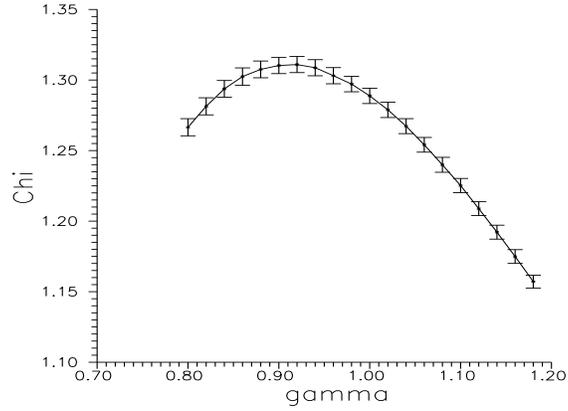,height=60mm,width=80mm,angle=0}
\vspace{1.5ex} \caption{\label{fig.6} Susceptibility $\chi = \langle H_Z^2
\rangle -
 \langle H_Z\rangle^2$ at $\beta = 15$ on the lattice $8^3\times 16$. }
\end{center}
\end{figure}

\begin{figure}
\begin{center}
 \epsfig{figure=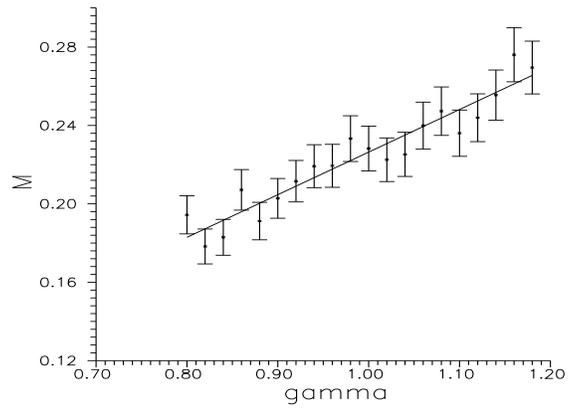,height=60mm,width=80mm,angle=0}
 \vspace{1.5ex}
\caption{\label{fig.3}  $M_Z$ as a function of $\gamma$.}
\end{center}
\end{figure}

\end{document}